\documentstyle[preprint,aps]{revtex}
\title{Induced quantum numbers in the \\ (2+1)-dimensional electron
gas\thanks{Work supported by the ``Deutsche Forschungsgemeinschaft''
under grant KL 256.}}
\author{Adrian Neagu\thanks{DAAD Fellow} and Adriaan M.J.
Schakel\thanks{Alexander von Humboldt Fellow}}
\address{Institut f\"{u}r Theoretische Physik, Freie Universit\"{a}t
Berlin, Arnimallee 14, 1000 Berlin 33 }
\begin{document}
\draft
\maketitle
\begin{abstract}
A gas of electrons confined to a plane
is examined in both the relativistic and nonrelativistic
case. Using a (0+1)-dimensional effective theory, a remarkably simple
method is proposed to calculate the spin density induced by an uniform
magnetic background field. The physical properties of possible fluxon
excitations are determined. It is found that while in the relativistic
case they can be considered as half-fermions (semions) in that they
carry half a fermion charge and half the spin of a
fermion, in the nonrelativistic case they should be thought of as
fermions, having the charge and spin of a fermion.
\end{abstract}
\section{Introduction}
Planar electron systems display peculiar phenomena which originate from
the Abelian nature of the rotation group SO(2) in two spatial
dimensions. Since the angular momentum is not quantized, quantum
statistics is allowed which continuously interpolate between
Bose-Einstein and Fermi-Dirac statistics \cite{W}. Particles obeying
such fractional statistics are called anyons \cite{Wb}. Their existence
in the context of the fractional quantum Hall effect (FQHE)
is generally accepted. The quasihole excitations of the Laughlin ground
state carry fractional charge and obey fractional statistics
\cite{L,ASW}.

An alternative approach to the FQHE due to Jain \cite{Jain} relates it
to the integer quantum Hall effect (IQHE). The basic concept of this
construction is that of a ``composite particle'', consisiting of an
electron bound to an even number of flux units. The IQHE of such
composite particles turns out to be equivalent to the FQHE of electrons.
All experimentally observed filling fractions are predicted in this way.
Also the observed hierarchy in stability of the various states is
naturally explained.

Whereas these systems involve nonrelativistic Landau levels,
relativistic levels, related to the Dirac Hamiltonian in an external
magnetic field, show up in a  certain type of doped two-dimensional
semimetals - materials with so-called diabolic points, where the
valence- and conduction bands intersect \cite{A,Gordon}.

These facts motivated us to study a planar gas of electrons occupying an
integer number of Landau levels in an uniform magnetic field, in both
the relativistic and nonrelativistic framework. In our treatment we
focus on induced quantum numbers like fermion charge and spin. Based on
our results, physical properties of possible fluxon excitations are
asessed. In 2+1 dimensions fluxons are point-like objects carrying one
magnetic flux unit $2\pi/e$, where $e$ is the electric charge of the
charge carriers in the system.  A fluxon may be pictured as the object
in the spatial plane that is obtained when this plane is pierced by a
magnetic flux tube.

In Sec.\ \ref{secrc} we consider the relativistic electron gas. We
extend a method recently proposed by one of us \cite{A} in order to
calculate the spin
induced by a magnetic background field with arbitrary strength, thus
generalizing the vacuum result of Paranjape \cite{Par}. We find a
close connection between induced fermion charge and induced spin,
reflecting the fact that spin and charge are not separated.
It is argued that a fluxon is, in fact, a half-fermion (semion)
having spin ${1 \over 4}$ and fermion charge ${1 \over 2}$. In the
nonrelativistic case, discussed in Sec.\ \ref{secnr}, a fluxon has
spin ${1 \over 2}$ and fermion charge $1$ and is, thus, a genuine
fermion. The close connection between induced spin and induced fermion
charge which we found in the relativistic case  is lost. In the last
section we explain that this is due to the fact that, contrary to the
relativistic case, in the nonrelativistic system the spin degree of
freedom is independent of the dynamics. We show that the induced spin in
the nonrelativistic electron gas is not related to a Chern-Simons term,
but to a so-called mixed Chern-Simons term, involving two different
gauge potentials, viz.\ the electromagnetic potential and one which
describes the spin degree of freedom.

\section{Relativistic computations} \label{secrc}
We consider a relativistic electron gas in two spatial dimensions in
the presence of an uniform magnetic field, as described by a massive
Dirac field at finite, positive chemical potential. In 2+1 dimensions
the Dirac algebra,
\begin{equation}
\{ \gamma ^{\mu} ,\gamma ^{\nu} \} = 2g^{\mu \nu } ,     \label{1}
\end{equation}
may be represented in terms of the Pauli matrices. We choose the
representation $\gamma ^{0}=\sigma ^{3} , \gamma ^{k}=i\sigma ^{k}$,
with $g^{\mu \nu }=g_{\mu \nu }={\rm diag}(1,-1,-1)$ the metric tensor
of Minkowski space.

The eigenvalues of the Dirac Hamiltonian play an important role
in our calculations. In order to make the discussion
self-contained, we provide a brief account following Johnson and Lippman
\cite{JL}. The Dirac equation reads
\begin{equation}
(i \gamma ^{\mu} D_{\mu}-m) \Psi = 0,                    \label{2}
\end{equation}
where $\Psi $ is a two-component Dirac spinor field with mass $m$, and
$D_{\mu}=\partial _{\mu }+ieA_{\mu }$ is the covariant derivative, assuring a
minimal coupling to the electromagnetic field with coupling constant $e$,
the electric charge. We describe the uniform magnetic field $B$ by the
vector potential $ A^{0}=A^{1}=0 ; A^{2}= B x^{1}$.
Separating the time variable by setting $\Psi (x^{0},
\mbox{\boldmath $x$})=\psi
(\mbox{\boldmath $x$}) \exp (-iEx^{0})$,
we write Eq.\ (\ref{2}) as an eigenvalue
equation for $ \psi (\mbox{\boldmath $x$})$:
\begin{equation}
(\mbox{\boldmath $\alpha$} \cdot
\mbox{\boldmath $\pi$} + \beta m) \psi = E \psi.       \label{3}
\end{equation}
The operator in parantheses is the Dirac Hamiltonian $ H_{{\rm D}}$,
it involves the matrices
$\alpha ^{k}=\gamma ^{0}\gamma ^{k},\ \beta =\gamma ^{0}$
and the gauge-invariant momentum $\pi ^{k}=iD^{k}$.
The idea is to look for the eigenvalues of the squared Hamiltonian
\begin{equation}
H_{{\rm D}}^{2}=\mbox{\boldmath $\pi$}^{2}-e B \sigma ^{3}+m^{2}  \label{4}
\end{equation}
in which one recognizes the Schr\"{o}dinger Hamiltonian $H_{\rm
S}$ of a spinless particle with mass ${1 \over 2}$ and charge $e$ in a
uniform magnetic field,
\begin{equation}
H_{{\rm S}}=\mbox{\boldmath $\pi$}^{2}.                           \label{5}
\end{equation}
This operator has the well-known oscillator eigenvalues
\begin{equation}
\epsilon _{n}= \Omega (n + {\scriptstyle \frac{1}{2}}) , \, \,\,
n=0,1,2\ldots,                                      \label{6}
\end{equation}
where $\Omega =2|e B |$ is the cyclotron frequency. The energy
eigenvalues corresponding to stationary solutions of the Dirac equation
(\ref{2}) now follow immediately. They are given by the relativistic
Landau levels \cite{JL}
\begin{equation}
E_{\pm n}=\pm \sqrt{m^{2}+2|e B |(n+{\scriptstyle \frac{1}{2}})
-2eB S_{\pm}},
                                                              \label{7}
\end{equation}
where the $+$ and $-$ signs correspond to positive and negative energy
spinors, respectively. The quantity $S_{\pm }$ denotes the eigenvalues
of the spin operator which in 2+1 dimensions is ${1 \over 2}\sigma
^{3}$; they are given by \cite{DJT,B1}
\begin{equation}
S_{\pm}=\pm \frac{1}{2} {\rm sgn}(m).                       \label{8}
\end{equation}
This can be easily shown by  observing that the energy eigenvalue
equation (\ref{3}) written in the rest frame,
\begin{equation}
\beta m \psi=\pm |m|\psi,                                   \label{rest}
\end{equation}
can be transformed into an eigenvalue equation for the spin
operator since $\beta =\sigma ^{3}$. This gives Eq.\ (\ref{8}) for the
spin of a particle in the rest frame. But the
spin is a pseudoscalar with respect to the Lorentz group [SO(2,1) in 2+1
dimensions], so a Lorentz-boost leaves it unchanged. We conclude that
the expression (\ref{8}) gives the spin of a particle in an arbitrary
frame.

The smallest energy eigenvalue is either $E_{+0} =+ \sqrt{m^{2}+|e
B |-2eB S_{+}}$, or $E_{-0} =- \sqrt{m^{2}+|e B |-2eB S_{-}}$; it has
the value $m \: {\rm sgn}(eB)$.

We next turn to the problem of the electron gas. Our starting point is
the Lagrangian
\begin{equation}
\mbox{$\cal L$} =\Psi ^{\dagger} (i\partial _{0}+\mu -H_{D})\Psi
+b \Psi ^{\dagger}\frac{\sigma ^{3}}{2}\Psi,  \label{10}
\end{equation}
where $\Psi $ is a Dirac spinor with two anticommuting (Grassmann)
components representing the positive and negative energy spinors, $\mu $
is a positive chemical potential which accounts for
the finite density, and $b$ is an external source which
couples to the spin density operator \({\scriptstyle \frac{1}{2}}\Psi
^{\dagger}\sigma ^{3}\Psi \). This last term enables us to compute the
induced spin density, i.e.\ the ground-state expectation value of the
spin density operator. It should be emphasized that $b$ has nothing to
do with the magnetic field $B$, i.e.\ the second term in (\ref{10}) is
not a Zeeman term which would appear in a nonrelativistic approximation.

Integrating out the fermionic degrees of freedom, one finds the one-loop
effective action:
\begin{equation}
S_{{\rm eff}}= \int d^3 x \, \mbox{$\cal L$}_{\rm eff} =
 -i\ln {\rm Det}(i\partial _{0}+\mu - H_{D} + \frac{b}{2} \sigma ^{3}),
\end{equation}
where Det stands for a functional determinant. Employing the
identity $\ln {\rm Det}={\rm Tr}\ln $, we obtain a functional trace,
which can be written in the energy representation as
\begin{eqnarray}
\mbox{$\cal L$}_{{\rm eff}}= \frac{|eB |}{2\pi } \sum_{n=0}^{\infty }
\int \frac{dk_{0}}{2 \pi i}
\bigl\{ && \ln [k_{0}+\mu - E_{+n}
+\frac{b}{2}  {\rm sgn}(m)] \nonumber \\ &&
+  \ln (k_{0}+\mu - E_{-n}- \frac{b}{2} {\rm
sgn}(m) ]\bigr\},                                            \label{14}
\end{eqnarray}
where $E_{\pm n}$ are the energy eigenvalues (\ref{7}), and $\pm {1
\over 2}{\rm sgn}(m)$ are the eigenvalues (\ref{8}) of the spin
operator.
We note that all the information about the system (except for the
degeneracy of a Landau level per unit area, $|eB |/2\pi $) is contained
in a (0+1)-dimensional theory, i.e.\ ordinary quantum mechanics \cite{A}.
This is due to the fact that the system is translation invariant (up to a
gauge transformation), so it suffice to study the system in a single
point.

In terms of the effective action one can express the ground state
expectation values of the fermion number density operator $\Psi
^{\dagger}\Psi$ and that of the spin density operator ${\scriptstyle
\frac{1}{2}} \Psi ^{\dagger} \sigma ^{3}\Psi$ as
\begin{equation}
\rho =\left. \frac{\partial \mbox{$\cal L$}_{\rm eff}}
{\partial \mu }\right|_{b=0}, \,\,\,\,\,
s=\left. \frac{\partial \mbox{$\cal L$}_{\rm eff}}
{\partial b}\right|_{b=0}.
                                                      \label{15}
\end{equation}
In this way we obtain from (\ref{14}) \cite{LSW,A}
\begin{equation}
\rho =\frac{|eB |}{2\pi }(N+ \frac{1}{2})\theta (\mu -|m|)
     -\frac{eB }{4\pi } {\rm sgn}(m) \theta (|m|-\mu )  \label{26}
\end{equation}
and
\begin{equation}
s = s_{\sim} + \frac{|eB |}{4\pi } {\rm sgn}(m) (N+ \frac{1}{2})\theta
(\mu -|m|)  -\frac{eB }{8\pi }\theta (|m|-\mu ),             \label{27}
\end{equation}
with $\theta$ the Heaviside unit step function. The integer $N$ denotes
the number of filled Landau levels
\begin{equation}
N= \left[ \frac{\mu ^{2}-m^{2}}{2|eB |} \right],
\end{equation}
where the integer-part function, $[x]$, denotes the largest integer less
than $x$. We assume that the value of the chemical potential does not
coincide with one of the Landau levels, thus avoiding the points in
which the integer-part function is discontinuous. These points
correspond to a partially filled Landau level. In deriving (\ref{26})
and (\ref{27}) we employed the integral
\begin{equation}
\int \frac{dk_{0} }{2\pi i}\ \frac{1}{k_{0} +\xi +
    ik_{0} \delta} =\frac{1}{2} {\rm sgn} (\xi ),     \label{19}
\end{equation}
where we have introduced the usual ``causal'' path-defining factor
$ik_{0}\delta$, with $\delta$ a small positive number. The first term in
expression (\ref{27}) for the induced spin density stands for the
infinite contribution stemming from negative energy states in the
Dirac sea,
\begin{equation}
s_{\sim} = -\frac{1}{2} \frac{|eB|}{2 \pi} {\rm sgn}(m)
\sum_{n=0}^{\infty} \theta (-E_{-n}).                      \label{dirac}
\end{equation}
A similar infinite term is not present in the expression for the induced
fermion number density. There, because of the spectral symmetry $E_{n+1}
= - E_{-n}$ $[{\rm sgn}(eBm)>0]$, or $E_{n-1} = -E_{-n}$ $[{\rm
sgn}(eBm)<0]$, only Landau levels with $|E_{\pm
n}|<\mu$ contribute. The contibutions to $\rho$ from levels outside this
energy interval cancel. We renormalize $s$ by subtracting the infinite
spin (\ref{dirac}) of the Dirac sea. This will be justified in a moment.
It then follows that the induced spin density is half the induced
fermion number density up to a sign ${\rm sgn}(m)$,
\begin{equation}
s=\frac{1}{2} {\rm sgn}(m) \rho .
\end{equation}
This result is reasonable. It shows that charge and spin are not
seperated, both are induced in a ratio that reflects the fact that these
quantum numbers are carried by a single particle, viz.\ the electron
with fermion charge 1 and spin ${1 \over 2}$.

The low-density limit, which corresponds to a chemical potential smaller
than the fermion mass $(\mu <|m|)$, deserves particular scrutiny. This
case is basically equivalent to vacuum ${\rm QED}_{2+1}$.
In this limit only the last term in (\ref{26}) survives, so that
the fermion number density induced into the vacuum by the
background magnetic field is \cite{NS,R,LSW,A}
\begin{equation}
\rho_{\rm vac}= -\frac{eB }{4\pi } {\rm sgn}(m),            \label{28}
\end{equation}
and, ignoring the contribution $s_{\sim}$ due to states inside the Dirac
sea,
\begin{equation}
s_{\rm vac}=-\frac{eB }{8\pi } = \frac{1}{2} {\rm sgn}(m) \rho_{\rm
vac}.                                                       \label{svac}
\end{equation}
The above results, which were derived for a constant background field,
also apply to cases where the magnetic field has a specific profile,
e.g.\ corresponding to a fluxon, the essential physics being captured by
the number of flux units that penetrate the spatial plane \cite{B2}.
The vacuum result (\ref{svac}) restricted to a single fluxon carrying one
magnetic flux unit $2\pi/e$ shows that it acquires fractional spin
$S_\otimes =- {1 \over 4}$. It was pointed out in Ref.\ \cite{GMW} that
this is in accord with the Chern-Simons term
which is generated at the quantum level when the system is placed in an
external {\it electro}magnetic field. This term is easily constructed
from (\ref{28}) by realizing that on account of Lorentz covariance the
induced fermion number current
density $\langle j^\mu \rangle$ in such a field,
described by the field strength $F^{\mu \nu}$, is
\begin{equation}
\langle j^\mu \rangle = \frac{e}{8 \pi} {\rm sgn}(m) \epsilon^{\mu \nu
\lambda} F_{\nu  \lambda}.
\end{equation}
This corresponds to a Chern-Simons term
\begin{equation}
\mbox{$\cal L$}_{\rm cs}= \frac{\theta e^2}{2}
  \epsilon_{\mu \nu \lambda} A^\mu \partial^\nu A^\lambda
                                                          \label{cs}
\end{equation}
in the effective Euler-Heisenberg Lagrangian, with $\theta=-{\rm sgn}(m)/(4
\pi)$. A Chern-Simons term imparts a spin
\begin{equation}
S_{\otimes} = {\rm sgn}(m)\pi \theta                     \label{display}
\end{equation}
to a fluxon. In our case this yields a spin $S_{\otimes}= -
{1 \over 4}$, in agreement with the result (\ref{svac}). One may
think of a fluxon as a half-fermion because it carries half the spin and
half the charge of a fermion.

To make connection with Paranjape's work \cite{Par}, who considered the
total angular momentum $J= S + L$ induced into the vacuum by $N_\phi$
fluxons, rather than the
induced spin $S = S_{\otimes} N_\Phi = -{1 \over 4} N_\Phi$, we note that
the exclusion principle forbids two anyons to be in the same angular
momentum state. So, when considering a state with $N_\phi$ semions,
these objects have to be put in succesive {\em orbital} angular momentum
states, and \cite{Forte}
\begin{equation}
L = 2 S_\otimes \sum_{n=1}^{N_\Phi} (n-1) = S_\otimes N_\phi (N_\phi -1).
                                                             \label{L}
\end{equation}
In this way, $J$ becomes
\begin{equation}
J = S+L = S_\otimes N^2_\Phi = - {1 \over 4} N^2_\Phi,
\end{equation}
which is, apart from a sign ${\rm sgn} (m)$, Paranjape's result
\cite{Par}.

{}From the full result (\ref{26}) for the induced fermion number density
we obtain the Chern-Simons coefficient
\begin{equation}
\theta = \frac{1}{2 \pi} {\rm sgn}(eB) (N + \frac{1}{2})
\theta (\mu - |m|) -
\frac{1}{4 \pi} {\rm sgn}(m) \theta (|m| - \mu),          \label{thfull}
\end{equation}
which, according to (\ref{display}) leads to a spin for a single fluxon
given by
\begin{equation}
S_\otimes = \frac{1}{2} {\rm sgn}(eBm) (N +
\frac{1}{2}) \theta(\mu - |m|) - \frac{1}{4} \theta(|m| - \mu).
\end{equation}
This yields, when multiplied with the density of fluxons, $eB/(2
\pi)$, the previous result (\ref{27}) with the contribution $s_\sim$ of
states in the Dirac sea omitted.

We next provide a further justification of our renormalization of the
expectation value of the spin
density operator, which consisted of subtracting the contributions
stemming from negative energy states in the Dirac sea. Since these
contributions are independent of the chemical potential, we may set
without loss of generality $\mu=0$ in our analysis. In this limit $s$
was given by  (\ref{svac}), implying a spin magnetic moment, or
magnetization $M$,
\begin{equation}
M = g_0 \mu_{\rm B} s = - \frac{e^2}{8 \pi |m|} B,     \label{mag}
\end{equation}
with $\mu_{\rm B}=e/(2|m|)$ the Bohr magneton and $g_0=2$ the electron
$g$-factor. The corresponding spin susceptibility, $\chi_{\rm P}$, is
\begin{equation}
\chi_{\rm P} = \frac{\partial M}{\partial B} =  - \frac{e^2}{8 \pi |m|},
                                                      \label{chi}
\end{equation}
where one should bear in mind that in (2+1)-dimensions $e^2$ has the
dimension of mass. We shall rederive this result, which hinges on our
premise that the first term $s_{\sim}$ in (\ref{27}) is to be omitted,
in an alternative way involving the ``proper-time'' regularization
developed by Schwinger \cite{Schwinger}.

To this end we carry out the $k_0$-integral in the effective Lagrangian
(\ref{14}) with $\mu=b=0$ to obtain
\begin{equation}
\mbox{$\cal L$}_{{\rm eff}}= \frac{|eB |}{2\pi } \frac{1}{2} \sum_n |E_{n}|,
\end{equation}
and introduce the ``proper-time'' representation of the square root
\cite{Hu,Cea}
\begin{equation}
\sqrt{a} = -\int_0^{\infty} \frac{d \tau}{(\pi \tau)^{1/2}} \frac{d}{d
\tau} \exp(-a\tau).                                           \label{sq}
\end{equation}
After a partial integration and after subtracting the $B$-independent
part, which corresponds to the free-electron contribution, one easily
finds
\begin{equation}
\mbox{$\cal L$}_{\rm eff} = - \frac{1}{8\pi^{3/2}} \int_0^{\infty} \frac{d
\tau}{\tau^{3/2}} {\rm e}^{-m^2\tau}\left(|eB| \frac{{\rm cosh}(2eB \sigma
\tau)}{{\rm sinh}(|eB|\tau)} - \frac{1}{\tau}\right).
                                                          \label{eff}
\end{equation}
Here, the sinh factor stems from the fact that the planar orbits of a
charged particle moving in a background magnetic field are quantized,
while the cosh factor, with $\sigma={1 \over 2}$, arises from the
magnetic moment and, thus, from the spin of the electrons. To obtain the
magnetic susceptibility we expand the effective Lagrangian (\ref{eff})
to second order in the magnetic field. This gives
\begin{equation}
{\cal L}_{{\rm eff},2} = {1 \over 2} \chi B^2,
\end{equation}
with $\chi$ the magnetic susceptibility
\begin{equation}
\chi = (-1)^{2 \sigma} \frac{e^2}{8\pi |m|} \left[ (2\sigma)^2 -
\frac{1}{3}\right] .                                    \label{chial}
\end{equation}
(We have written this formula in a general form valid for spin $\sigma = 0,
{1 \over 2},1.$)
The first term, with $\sigma={1 \over 2}$, is the spin contribution which
precisely yields the previous result (\ref{chi}). This justifies the
renormalization procedure we adopted.

Incidentally, it follows from (\ref{chial}) that for relativistic spin-${1
\over 2}$ particles in two (and also in three) space dimensions the
spin contribution is three times as large as the orbital contribution.
The same ratio is found for a nonrelativistic electron gas at small
magnetic fields in three space dimensions, where
\begin{equation}
\chi = (-1)^{2\sigma+1}  2 \mu_{\rm B}^2 \, \nu_{3D}(0) \left[(2\sigma)^2 -
\frac{1}{3}\right],                                \label{chi3D}
\end{equation}
with $\nu_{3D}(0) = m k_F/2\pi^2$ the three-dimenional density of
states per spin degree of freedom at the Fermi sphere. However, whereas
usually the spin contribution is paramagnetic $(\chi>0)$ and the orbital
contribution is diamagnetic $(\chi<0)$, Eq.\ (\ref{chial}) reveals
exactly the  opposite behaviour. Instead of screening the external
field the planar motion of relativistic electrons is such as to enhance
the field. Since the diamagnetic (sic!) spin contribution dominates,
the overall effect in vacuum QED$_{2+1}$ is nevertheless a screening of
external fields $(\chi<0)$.
\section{Nonrelativistic calculations}    \label{secnr}
In this section we treat a nonrelativistic electron gas confined to a
plane. We expect that some new qualitative features arise from the fact
that in this case the spin degree of freedom is not enslaved by the
dynamics. We continue to use a relativistic notation with $\partial_\mu
=(\partial_0, \mbox{\boldmath $\nabla$}$,
 $\partial ^\mu =(\partial_0, -\mbox{\boldmath $\nabla$})$, where
$\mbox{\boldmath $\nabla$}$ is the gradient operator,
and $A^\mu = (A^0, \mbox{\boldmath $A$})$.

Let us consider the Lagrangian
\begin{equation}
\mbox{$\cal L$} =\Psi ^{\dagger}(i\partial _{0}+\mu -H_{{\rm P}})\Psi +
      b \Psi ^{\dagger}\frac{\sigma ^{3}}{2}\Psi          \label{30}
\end{equation}
which governs the dynamics of the Pauli spinor field $\Psi$,
with Grassmann components $\psi _{\uparrow}$ and
$\psi _{\downarrow}$ describing the electrons with spin-$\uparrow$
and $\downarrow$. The role of the chemical potential $\mu$ and the
spin source $b$ is the same as in the previous calculation. The Pauli
Hamiltonian
\begin{equation}
H_{{\rm P}}=\frac{1}{2m}(i\mbox{\boldmath $\nabla$}
+e\mbox{\boldmath $A$})^{2}- g_0 \mu_{\rm B}
\frac{\sigma^3}{2} B + eA_{0},                              \label{31}
\end{equation}
with $\mu_{\rm B}=e/2m$ the Bohr magneton and $g_0$ the electron
$g$-factor,
contains a Zeeman term which couples the electron spins to the
background magnetic field. Usually this term is omitted. The reason is
that in realistic systems the $g$-factor is much larger than two, the
value for a free electron. In strong magnetic fields relevant to the
QHE the energy levels of spin-$\downarrow$ electrons are
too high and cannot be occupied; the system is spin
polarized, and the electron spin is irrelevant to the problem.
Setting again $A^{0}=A^{1}=0,\ A^{2}=B x^{1}$, one finds as eigenvalues
for $H_{{\rm P}}$
\begin{equation}
E_{n,\pm} = \frac{|eB|}{m} (n+{1 \over 2}) - \frac{eB}{m} S_\pm,
                                                         \label{nrll}
\end{equation}
with $S_\pm = \pm {1 \over 2}$ for spin-$\uparrow$ and spin-$\downarrow$
electrons, respectively. We note that in the nonrelativistic limit,
corresponding to taking $m \rightarrow + \infty$, the relativistic
Landau levels (\ref{7}) reduce to
\begin{equation}
E_{+n} \rightarrow {\rm const.} + \frac{|eB|}{m} (n+{1 \over 2}) -
\frac{eB}{2m},
\end{equation}
where we omitted the negative energy levels which have no meaning in
this limit. The main difference with (\ref{nrll}) stems from the fact
that there the spin degree of freedom is considered as an independent
quantity, not enslaved by the dynamics as is the case in the
relativistic problem.

The induced fermion number density and spin density may be obtained in a
similar calculation as in the preceding section. From the effective
action,
\begin{equation}
S_{{\rm eff}}=-i{\rm Tr}\ln(i\partial _{0}-H_{P}+\mu +\frac{b}{2}\sigma
^{3}),
\end{equation}
we obtain
\begin{equation}
\mbox{$\cal L$}_{{\rm eff}}=\frac{|eB |}{2\pi } \sum_{n=0}^{\infty }
\int \frac{dk_{0}}{2 \pi i} \left[ \ln (k_{0}-E_{n,+}+\mu +\frac{b}{2})+
\ln (k_{0}-E_{n,-}+\mu -\frac{b}{2})\right].            \label{nreff}
\end{equation}
The only difference with the relativistic computation is that instead of
integrals of the type (\ref{19}), we now encounter integrals of the form
\begin{equation}
\int \frac{dk_{0}}{2\pi i}\: \frac{e^{ik_{0}\delta }}
{k_{0} +\xi +ik_{0}\delta } = \theta (\xi )               \label{37}
\end{equation}
containing, as usual in nonrelativistic calculations \cite{FW}, an
additional convergence term $\exp (ik_{0}\delta)$.
The resulting value of the induced fermion number density is
\begin{equation}
\rho = \frac{|eB|}{2\pi} (N_+ + N_-),                       \label{39}
\end{equation}
with $N_\pm$ the number of filled Landau levels for spin-$\uparrow$ and
spin-$\downarrow$ electrons,
\begin{equation}
N_\pm = \left[ \frac{m \mu_\pm }{|eB|} + \frac{1}{2}\right] ,
                                                      \label{npm}
\end{equation}
and
\begin{equation}
\mu_\pm = \mu + \frac{eB}{m} S_\pm                     \label{chemical}
\end{equation}
their effective chemical potentials.
The square brackets denote again the integer-part function.
Implicit in this framework is the assumption that, just like in the
relativistic case, the chemical potential lies between two Landau levels.
The induced fermion number density (\ref{39}) is related to a
Chern-Simons term (\ref{cs}) in the effective action, with
\begin{equation}
\theta = {\rm  sgn}(eB) \frac{1}{2\pi} (N_+ + N_-).      \label{nrth}
\end{equation}
Because of the presence of the ${\rm  sgn}(eB)$ factor, which changes
sign under a parity transformation, this Chern-Simons term is invariant
under such transformations. The induced spin density turns out to be
independent of $N_\pm$, viz.\
\begin{equation}
s=\frac{eB }{4\pi}.                                       \label{41}
\end{equation}
This follows from the symmetry in the spectrum $E_{n+1,+} =
E_{n,-}$ $(eB>0)$, or $E_{n,+} = E_{n+1,-}$
$(eB<0)$. The magnetic moment, $M$, is according to (\ref{mag}) obtained
from (\ref{41}) by multiplying $s$ with twice the Bohr magneton,
$\mu_{\rm B}$. This leads to the text-book result for the magnetic spin
susceptibity $\chi_{\rm P}$
\begin{equation}
\chi_{\rm P} = \frac{\partial M}{\partial B} = \frac{e^2}{4 \pi m} = 2 \mu_{\rm
B}^2 \, \nu_{2D}(0),                                  \label{textb}
\end{equation}
with $\nu_{2D}(0) = m/(2\pi)$ the density of states per spin degree of
freedom in two space dimensions.

At zero field, $\rho$ reduces to the standard fermion number density in
two space dimensions
$\rho \rightarrow \mu m/\pi = k_{\rm F}^2/(2\pi)$,
where $k_{{\rm F}}$ denotes the Fermi momentum. A single fluxon carries
according to (\ref{41}) a spin $S_\otimes = {1 \over 2}$ and, since for
small fields
\begin{equation}
\rho \rightarrow \frac{\mu m}{\pi } + \frac{|eB|}{2 \pi},
\end{equation}
also one unit of fermion charge. That is, in the
nonrelativistic electron gas the fluxon may be thought of as a
fermion in that it has both the spin and charge of a fermion.
However, the close connection between spin of a fluxon and induced
Chern-Simons term for arbitrary fields that we found in the relativistic
case is lost. This can be traced back to the fact that in the
nonrelativistic case the electron spin is an independent degree of
freedom. In the next section we point out that the spin of the fluxon
does not derive from the ordinary Chern-Simons term, but from a
so-called mixed Chern-Simons term. Such a term is absent in the
relativistic case.

To see how the spin contribution (\ref{textb}) to the magnetic
susceptibility compares to the orbital contribution we evaluate the
$k_0$-integral in the effective action (\ref{nreff}) with $b=0$ to
obtain
\begin{equation}
\mbox{$\cal L$}_{\rm eff} =  \frac{|eB|}{2 \pi} \sum_{n=0}^{\infty}
\sum_{\varsigma =\pm} (\mu - E_{n,\varsigma}) \theta(\mu -
E_{n,\varsigma}).
\end{equation}
The summation over $n$ is easily carried out with the result for small
fields
\begin{equation}
\mbox{$\cal L$}_{\rm eff} = \frac{1}{4 \pi} \sum_{\varsigma =\pm} \left[
\mu_\varsigma^2 m - \frac{(eB)^2}{4 m} \right]
              = \frac{\mu^2 m}{2 \pi} + \frac{(eB)^2}{8 \pi m}
[(2\sigma)^2 -1],                                    \label{nr}
\end{equation}
where $\sigma={1 \over 2}$ and  $\mu_\pm$ is given by (\ref{chemical}).
The first term in the right-hand side of (\ref{nr}), which is
independent of the magnetic field, is the free particle contribution
\begin{equation}
\frac{\mu^2 m}{2 \pi} = -2 \int \frac{d^2 k}{(2\pi)^2} \left(
\frac{k^2}{2m} - \mu \right) \theta \left(\mu - \frac{k^2}{2m} \right).
\end{equation}
The second term yields the low-field susceptibility
\begin{equation}
\chi = (-1)^{2 \sigma +1}  2 \mu_{\rm B}^2 \, \nu_{2D}(0) \left[(2\sigma)^2 -
1)\right].                                            \label{nrchi}
\end{equation}
Equation (\ref{nrchi}) shows that
the ratio of orbital to spin contribution to $\chi$ is different
from the three-dimensional case (\ref{chi3D}). Also, whereas a $3D$
electron gas is paramagnetic $(\chi>0)$ because of the dominance of the
spin contribution, the $2D$ gas is  not magnetic $(\chi=0)$ at small
fields since the orbital and spin contributions to $\chi$ cancel.
\section{Mixed Chern-Simons term}
In this section we investigate the origin of the induced spin density
(\ref{41}) we found in the nonrelativistic electron gas. To this end we
slightly generalize the theory (\ref{30}) and consider the Lagrangian
\begin{equation}
\mbox{$\cal L$} =\Psi ^{\dagger}\left[ i\partial _{0} - e A_0 + \mu
-\frac{1}{2m}(i\mbox{\boldmath $\nabla$} + e \mbox{\boldmath $A$})^{2}
\right] \Psi +\frac{e}{m}B^{a}\Psi ^{\dagger }\frac{\sigma ^{a}}{2}\Psi.
                                                             \label{43}
\end{equation}
It differs from (\ref{30}) in that the spin source term is omitted, and
in that the magnetic field in the Zeeman term is allowed to point in any
direction in some internal space labelled by latin indices
$a,b,c=1,2,3$. As a result also the spin will have components in this
space. It is convenient to consider a magnetic field whose
direction in the internal space varies in space-time. We set
\begin{equation}
B^{a}(x)=B n^{a}(x),                                       \label{n}
\end{equation}
with $n^a$ a unit vector in the internal space. The gauge potential
$A_\mu$ appearing in the first term of (\ref{43}) still gives $\epsilon_{i j}
\partial_i A^j = B$. Equation (\ref{n}) allows us to make the decomposition
\begin{equation}
\Psi (x)=S(x)\chi (x) \; ; \; \; S^{\dagger }S=1,
\end{equation}
with $S(x)$ a local SU(2) matrix which satisfies
\begin{equation}
\mbox{\boldmath $\sigma$} \cdot  \mbox{\boldmath $n$}(x)=
S(x)\sigma ^{3}S^{\dagger }(x).
                                                        \label{46}
\end{equation}
In terms of these new variables the Lagrangian (\ref{43}) becomes
\begin{equation}
\mbox{$\cal L$} =\chi ^{\dagger }\left[ i\partial _{0} - eA_0 -V_{0}
    + \mu - \frac{1}{2m}
    (i\mbox{\boldmath $\nabla$} + e \mbox{\boldmath $A$}
    + \mbox{\boldmath $V$})^{2}\right]\chi +
     \frac{eB }{2m}\chi ^{\dagger}\sigma ^{3}\chi,           \label{47}
\end{equation}
where the $2\times 2$ matrix $V_{\mu }=-iS^{\dagger}(\partial _{\mu}S)$
represents an element of the su(2) algebra, which can be written in
terms of (twice) the generators  $\sigma^a$ as
\begin{equation}
V_{\mu }=V_{\mu }^{a} \sigma ^{a}.
\end{equation}
In this way the theory takes formally the form of a gauge theory with
gauge potential $V_{\mu }^{a}$. In terms of the new fields the spin
density operator,
\begin{equation}
j_{0}^{a}=\Psi ^{\dagger}\frac{\sigma ^{a}}{2}\Psi,            \label{49}
\end{equation}
becomes \cite{Ko}
\begin{equation}
j_{0}^{a}=R_{ab}\chi ^{\dagger}\frac{\sigma ^{b}}{2}\chi = - \frac{1}{2}
R_{ab} \frac{\partial \mbox{$\cal L$} }{\partial V_{0}^{b}}.   \label{50}
\end{equation}
In deriving the first equation we employed the identity
\begin{equation}
S^{\dagger}(\mbox{\boldmath $\theta$})\sigma ^{a}S
(\mbox{\boldmath $\theta$})=
   R_{ab}(\mbox{\boldmath $\theta$})\sigma ^{b},              \label{51}
\end{equation}
which relates the SU(2) matrices in the $j={1 \over 2}$
representation, $S(\mbox{\boldmath $\theta$})=
\exp({i \over 2} \mbox{\boldmath $\theta$} \cdot
\mbox{\boldmath $\sigma$})$, to those in the adjoint representation $(j=1)$,
$R(\mbox{\boldmath $\theta$})=\exp(i\mbox{\boldmath $\theta$}
\cdot \mbox{\boldmath $ J^{\rm adj}$})$.
The matrix elements of the generators in the latter representation are
$\left(J_{a}^{{\rm adj}}\right)_{bc}=-i\epsilon _{abc}$.

The projection of the spin density $j_0^a$ onto the spin quantization
axis, i.e.\ the direction $n^a$ of the applied magnetic field \cite{Ko},
\begin{equation}
\mbox{\boldmath $n$} \cdot \mbox{\boldmath $j$}_{0}=
 -\frac{1}{2} \frac{\partial \mbox{$\cal L$}}
     {\partial V_{0}^{3}},                                        \label{54}
\end{equation}
only involves the spin gauge field $V_\mu^3$. So when calculating the
induced spin density $s = \langle \mbox{\boldmath $n$}
\cdot \mbox{\boldmath $j$}_{0} \rangle$ we
may set the fields $V^1_\mu$ and $V^2_\mu$ equal to zero and consider the
simpler theory
\begin{equation}
\mbox{$\cal L$} = \sum_{\varsigma = \pm} \chi_{\varsigma}^{\dagger }
\left[ i\partial _{0}
- eA_0^{\varsigma} + \mu_{\varsigma} -
\frac{1}{2m} (i \mbox{\boldmath $\nabla$} + e
\mbox{\boldmath $A$}^{\varsigma})^{2}\right]\chi_{\varsigma},
\end{equation}
where the effective chemical potentials for the spin-$\uparrow$ and
spin-$\downarrow$ electrons are given in (\ref{chemical})
and $eA_\mu^\pm = eA_\mu \pm V_\mu^3$. Both components
$\chi_\uparrow$ and $\chi_\downarrow$ induce a Chern-Simons term, so
that in total we have
\begin{eqnarray}
\mbox{$\cal L$}_{\rm cs} &=& \frac{e^2}{2} \epsilon^{\mu \nu \lambda}
( \theta_+ A^+_\mu \partial_\nu
A^+_\lambda + \theta_- A^-_\mu \partial_\nu  A^-_\lambda) \label{total}  \\
           &=&  \frac{\theta_+ + \theta_-}{2} \epsilon^{\mu \nu \lambda} (e^2
A_\mu \partial_\nu  A_\lambda + V^3_\mu \partial_\nu  V^3_\lambda)
+ e (\theta_+ - \theta_-)
\epsilon^{\mu \nu \lambda} V^3_\mu \partial_\nu  A_\lambda,  \nonumber
\end{eqnarray}
where the last term involving two different vector potentials is a mixed
Chern-Simons term. The coefficients are given by
\begin{equation}
\theta_\pm = \frac{1}{2 \pi} {\rm sgn}(eB) N_\pm ,
\end{equation}
assuming that $|eB|>{1 \over 2} |\epsilon_{i j} \partial_i V_j^3|$, so
that the sign of $eB$ is not changed by spin gauge contributions. The
integers $N_\pm$ are the number of filled Landau levels for
spin-$\uparrow$ and spin-$\downarrow$ electrons given by (\ref{npm}).
Since $N_+ -N_- = {\rm sgn} (eB)$, we obtain for the induced spin
density $s$ precisely the result (\ref{41}) we found in the preceeding
section,
\begin{equation}
s = \langle \mbox{\boldmath $n$} \cdot \mbox{\boldmath $j$}_{0}
\rangle = -\frac{1}{2} \left.
\frac{\partial \mbox{$\cal L$}_{\rm eff}}
{\partial V_{0}^{3}} \right|_{V_\mu^3 = 0} =
\frac{eB}{4 \pi}.
\end{equation}
The present derivation clearly shows that the induced spin in
the nonrelativistic electron gas originates not from the standard
Chern-Simons term (\ref{cs}), but from the mixed Chern-Simons term
involving the electromagnetic and spin gauge potential.

The first term in (\ref{total}) is a standard Chern-Simons term, the
combination $\theta_+ + \theta_-$ precisely reproduces the result
(\ref{nrth}) and is related to the induced fermion number density
(\ref{39}).
\section*{Acknowledgements}
A.N.\ gratefully acknowledges financial support from the ``Deutscher
Akademischer Austauschdienst'' (DAAD), the work of A.M.J.S.\ is
financially supported by the Alexander von Humboldt Foundation.


\begin{references}
\bibitem{W} F. Wilczek, {\em Phys.\ Rev.\ Lett.\ } {\bf 48} (1982) 1144;
                        {\em Phys.\ Rev.\ Lett.\ } {\bf 49} (1982) 957.
\bibitem{Wb} F. Wilczek, {\em Fractional Statistics and Anyon
            Superconductivity} (World Scientific, 1990).
\bibitem{L} R. B. Laughlin, {\em Phys.\ Rev.\ Lett.\ } {\bf 50}
                             (1983) 1395.
\bibitem{ASW} D. Arovas, J. R. Schrieffer, and F. Wilczek,
              {\em Phys.\ Rev.\ Lett.\ } {\bf 53} (1984) 722.
\bibitem{Jain} J. K. Jain, {\em Phys.\ Rev.\ Lett.\ } {\bf 63} (1989)
199; {\em Adv.\ Phys.} {\bf 41} (1992) 105.
\bibitem{A} A. M. J. Schakel, {\em Phys.\ Rev.\ } {\bf D43} (1991) 1428.
\bibitem{Gordon} A. M. J. Schakel and G. W. Semenoff, {\em Phys.\ Rev.\
Lett.} {\bf 66} (1991) 2653.
\bibitem{Par} M. B. Paranjape, {\em Phys. Rev. Lett.} {\bf 55} (1985) 2390;
              {\it ibid.} {\bf 57} (1986) 500.
\bibitem{JL} M. H. Johnson and B. A. Lippmann, {\em Phys.\ Rev.\ }
              {\bf 76} (1949) 828.
\bibitem{DJT} S. Deser, R. Jackiw, and R. Templeton, {\em Phys.\ Rev.\
              Lett.\ } {\bf 48} (1982) 975; {\em Ann.\ Phys.}
              (N.Y.){\bf 140} (1982) 372.
\bibitem{B1} D. Boyanovsky, R. Blankenbecler, and R. Yahalom,
             {\em Nucl.\ Phys.\ } {\bf B270} (1986) 483.
\bibitem{LSW} J. D. Lykken, J. Sonnenschein, and N. Weiss,
              {\em Phys.\ Rev.\ } {\bf D42} (1990) 2161;
              {\em Int.\ J.\ Mod.\ Phys.\ } {\bf A6} (1991) 1335.
\bibitem{NS} A. J. Niemi and G. W. Semenoff, {\em Phys.\ Rev.\ Lett.\ }
            {\bf 51} (1983) 2077.
\bibitem{R} A. N. Redlich, {\em Phys.\ Rev.\ Lett.\ } {\bf 52} (1984) 18;
                           {\em Phys.\ Rev.\ } {\bf D29} (1984) 2366.
\bibitem{B2} R. Blankenbecler and D. Boyanovsky,
             {\em Phys.\ Rev. } {\bf D34} (1986) 612.
\bibitem{GMW} A. S. Golhaber, R. Mackenzie, and F. Wilczek, {\em Mod.\
Phys.\ Lett.} {\bf A4} (1989) 21.
\bibitem{Forte} L. S. Forte, {\em Rev. Mod. Phys.} {\bf 64} (1992) 193.
\bibitem{Schwinger} J. Schwinger, {\em Phys. Rev.} {\bf 82} (1951) 664.
\bibitem{Hu} R. J. Hughes, {\em Phys. Lett.} {\bf 148B} (1984) 215.
\bibitem{Cea} P. Cea, {\em Phys. Rev.} {\bf D32} (1985) 2785;  We disagree
with the computation of the vacuum energy in this paper.
\bibitem{FW} A. L. Fetter and J. D. Walecka, {\em Quantum Theory of
             Many-Body Systems} (Pergamon, Oxford, 1967).
\bibitem{Ko} A. M. J. Schakel, in {\em Proceedings of the K\"orber
Symposium on Superfluid $^3$He in Rotation}, Helsinki, 1991, edited by
M. M. Salomaa, {\em Physica} {\bf B178} (1992) 280.
\end{references}
\end{document}